\documentclass[twocolumn]{revtex4}
\usepackage{hyperref}
\usepackage{amsmath}
\usepackage{textcomp}
\usepackage{units}
\usepackage{graphicx}

\usepackage{amssymb}

\begin{document}

\title{Staging of RF-accelerating units in a MEMS-based ion accelerator}

\author{A. Persaud}
\email[E-mail:]{apersaud@lbl.gov}
\author{P.~A. Seidl}
\author{Q. Ji}
\author{E. Feinberg\footnote{Now XTD-IDA, Los Alamos National Laboratory, Los Alamos, NM 87545}}
\author{W.~L. Waldron}
\author{T. Schenkel}
\affiliation{Accelerator Technology \& Applied Physics, E. O. Lawrence Berkeley National Laboratory\\ 1 Cyclotron Road, Berkeley CA 94720, United States}

\author{S. Ardanuc}
\author{K.~B. Vinayakumar}
\author{A. Lal}
\affiliation{School of Electrical and Computer Engineering, Cornell University\\ 120 Phillips Hall, Ithaca NY 14853, United States}

\begin{abstract}
  Multiple Electrostatic Quadrupole Array Linear Accelerators
  (MEQALACs) provide an opportunity to realize compact radio-frequency
  (RF) accelerator structures that can deliver very high beam
  currents. MEQALACs have been previously realized with acceleration
  gap distances and beam aperture sizes of the order of
  centimeters. Through advances in Micro-Electro-Mechanical Systems
  (MEMS) fabrication, MEQALACs can now be scaled down to the
  sub-millimeter regime and batch processed on wafer substrates. In
  this paper we show first results from using three RF stages in a
  compact MEMS-based ion accelerator. The results presented show
  proof-of-concept with accelerator structures formed from printed
  circuit boards using a $3\times 3$ beamlet arrangement and noble gas
  ions at \unit[10]{keV}. We present a simple model to describe the
  measured results. We also discuss some of the scaling behaviour of a
  compact MEQALAC. The MEMS-based approach enables a low-cost, highly
  versatile accelerator covering a wide range of currents
  ($\unit[10]{\mu A}$ to $\unit[100]{mA}$) and beam energies
  (\unit[100]{keV} to several \unit{MeV}). Applications include
  ion-beam analysis, mass spectrometry, materials processing, and at
  very high beam powers, plasma heating.
\end{abstract}
\maketitle

\section{Introduction}
A driving force in the development of new accelerators for
applications in research and industry is reducing the cost and the
size of the instrument while at the same time achieving higher beam
intensity. In \citet{Persaud2016}, we proposed a new concept to
achieve high beam intensities using a very compact multi-beamlet
accelerator structure. The concept is based on earlier work on
Multiple Electrostatic Quadrupole Array Linear Accelerators (MEQALACs)
in the 1980s by \citet{Maschke1979} and has been implemented by, for
example, \citet{Urbanus1989} using beamlet apertures on the order of
centimeters. We propose to reduce the size by greater than an order of
magnitude by implementing the MEQALAC concept using
micro-electro-mechanical systems (MEMS) structures, reducing the
aperture size to the (sub-) millimeter scale and possibly to the scale of tens
of micrometers. The MEQALAC is based on the fact that decreasing the
beam aperture size in the electrostatic quadrupoles (ESQs), which
supply the needed beam focusing, leads to higher transportable beam
current densities. Therefore, it is advantageous to replace a large
aperture and its focusing quadrupoles with multiple small apertures
and quadrupoles closely packed in the transverse plane. The smallest
achievable size is determined by fabrication errors. For example,
fabrication errors will lead to the displacement of quadrupole centers
from the desired beamline axis, leading to beam centroid oscillations
and particle loss.  Using MEMS technology will push these fabrication
errors into the micrometer range and therefore allow better alignment
and high beam intensities. Using multiple apertures in parallel will
then allow to increase the total transported beam current compared to
single aperture accelerators.

The accelerator structure contains two elements: electrostatic
quadrupole lenses to focus the beam and transport it along the beam
line, and radio-frequency (RF) acceleration. Compared to non-RF
accelerators, one of the advantages of a MEQALAC structure is the use
of relatively small voltages ($<\unit[10]{kV}$) to achieve high beam
energies (up to \unit{MeV}). All elements are to be implemented in a
silicon wafer structure, so that they can be easily fabricated using
MEMS technology. For the initial proof-of-concepts experiment, both RF
and ESQ structures have been implemented in circuit boards (FR-4)
using millimeter-scale structures and a $3\times 3$ array of
beamlets. The FR-4 boards are fabricated by laser cutting, see
\citet{Persaud2016} for details on the fabrication procedure. In this
paper we will focus on experiments using two and three RF units that
have been realized in FR-4. We will show results from staging these RF
units and compare these measurements with a simple 1D-model.

\section{Experimental Setup}
\begin{figure}[t]\vspace*{4pt}
\centering
\includegraphics[width=\linewidth]{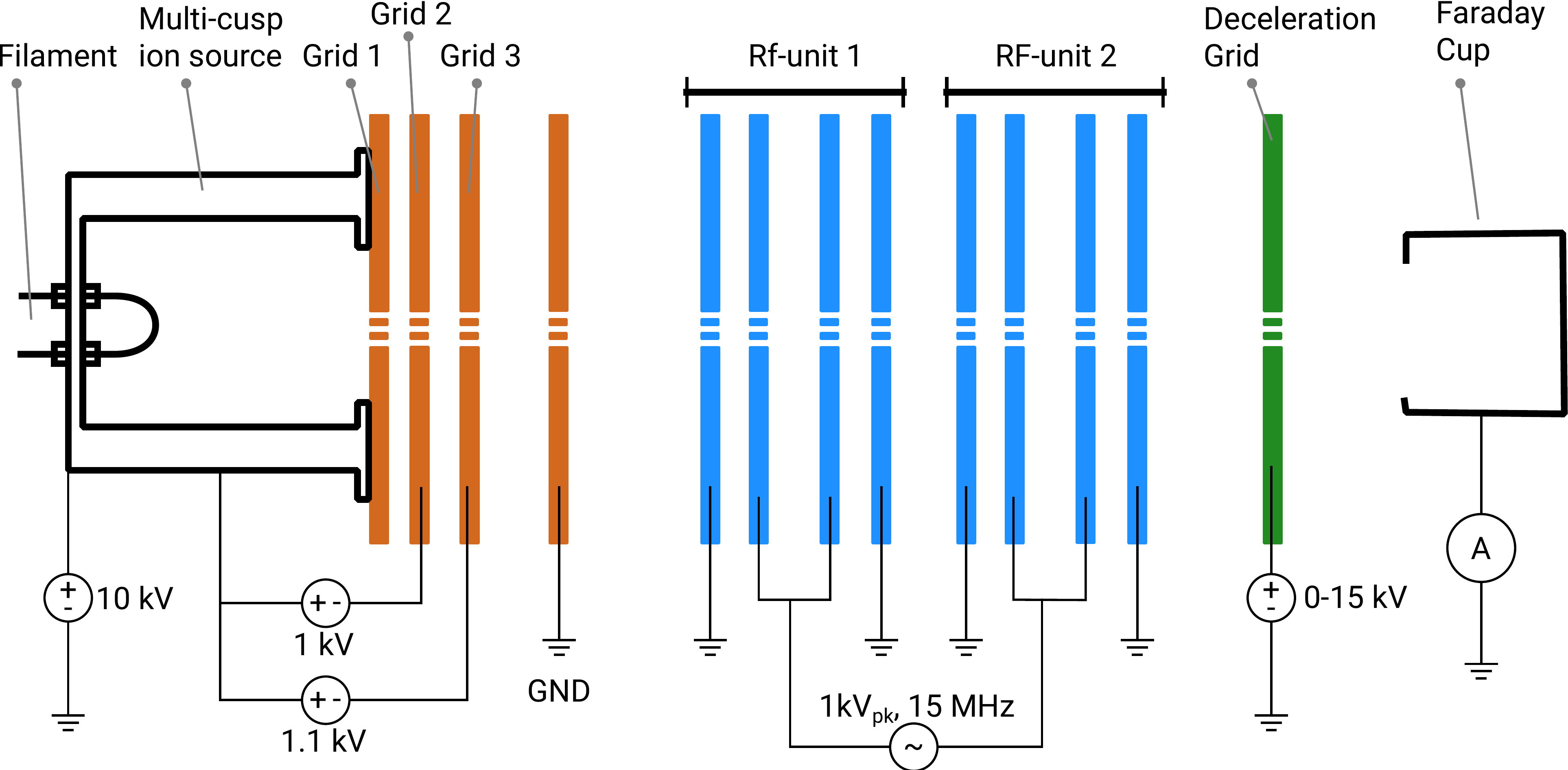}
\caption{The experimental setup consists of an ion source at high
  potential, an extraction grid system, the RF units, a
  retarding-potential analyzer, and a Faraday cup.}
\label{fig:setup}
\end{figure}

\begin{figure}[t]\vspace*{4pt}
\centering
\includegraphics[angle=270,width=0.8\linewidth]{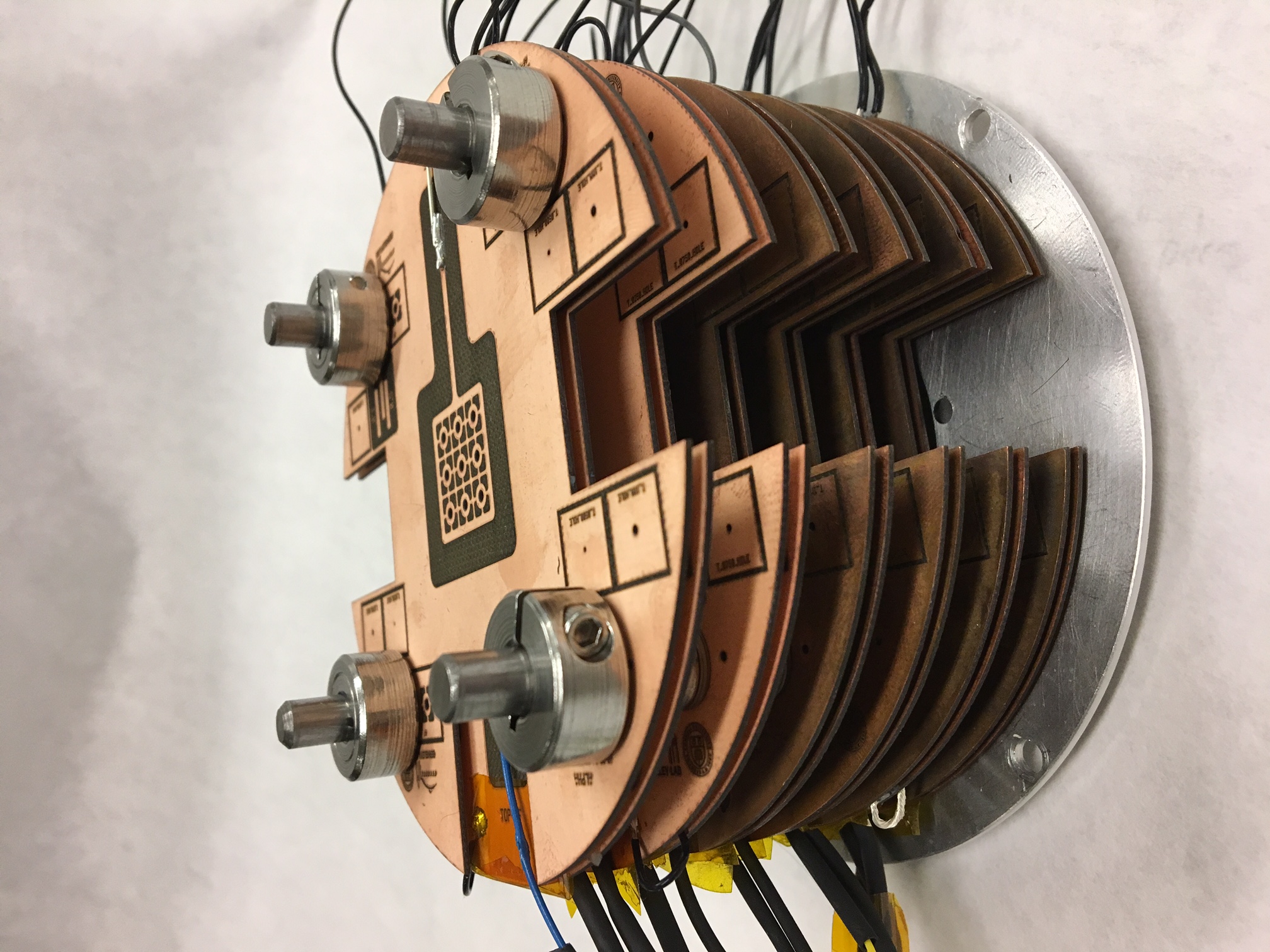}
\caption{Assembly of three RF units with drift spaces between the
  acceleration gaps set up for argon ions at \unit[11]{keV} kinetic energy and
  an RF frequency of \unit[15]{MHz} at \unit[1]{kV} amplitude.}
\label{fig:assembly}
\end{figure}

To demonstrate the staging of several RF units, the setup shown in
Fig.~\ref{fig:setup} was used.
A multi-cusp ion source driven by a filament discharge, as described
in \citet{Ji2016}, is used to create the injected ion beam. The flow
of argon gas is set to achieve a pressure of \unit[7]{mTorr} inside
the source body. The filament operates at \unit[3.5]{V} and
\unit[36]{A} for 7.5 seconds. After 6 seconds of filament heating an
arc-pulse of \unit[-100]{V} is applied for $\unit[300]{\mu s}$ between
the filament and the source housing to ignite the plasma. The ion
beams (a $3\times 3$ array with a pitch of 5 mm) are then extracted
between two aligned hole plates (grids 1\&2 in Fig.~\ref{fig:setup})
by floating grid 1 and biasing grid 2 to \unit[-1]{kV}. A third hole
plate (grid 3 in Fig.~\ref{fig:setup}) is available at the source to
enable fast beam pulsing, but this feature was not utilized in the
experiments described in this paper and the plate was instead biased
at a constant \unit[-1.1]{kV}. To be able to operate the RF and
quadrupole wafers at ground potential, the ion source and the source
grid power supplies are floating on high voltage
($<\unit[12]{kV}$). This voltage will accelerate the ions to a
grounded electrode that marks the end of the source setup. Afterwards
the ions enter two or more RF units. A single RF unit consists of four
wafers. The outer wafers are grounded and RF is applied to the inner
wafers, so that for each beamlet a field between wafer 1\&2 and 3\&4
is created, but the region between wafer 2\&3 is kept field free. Ions
are accelerated or decelerated when entering the regions with electric
fields. The length of the field free region can be chosen so that ions
are accelerated in both gaps if they arrive at the right phase of the
RF signal. For this to happen the RF needs to change sign during the
time the ion drifts through the field free region. This condition is
met when the drift length is $\frac{\beta \lambda}{2}$ where $\beta$
is the velocity of the ion relative to the speed of light and
$\lambda$ is the wavelength of the RF signal. RF units can be stacked
by adding another drift region between neighboring RF units. As the
ions gain energy along the accelerator structure, the length of the
drift regions has to increase along the beamline, see
Fig.~\ref{fig:assembly} for an image of an assembly of three RF units
as used in the experiments.
To measure the energy distribution
of the ions a retarding-field grid is positioned between the exit of
the last RF unit and a Faraday cup to measure the beam current. By
increasing the retarding potential on the grid, ions with a lower beam
energy than the applied grid voltage will be deflected and cannot
enter the Faraday cup. A large opening at the Faraday cup ensures that
all ions passing through the retarding grid are collected.

In our setup, we generate the RF signal from a RF-signal generator
(Keysight 33500B) that gets amplified to $\unit[<1]{W}$ signal using a
Mini-Circuits ZHL-2-8 and in turn drives the gate of an RF-MOSFET to
excite a tuned resonator. The main RF loads are in the MOSFET circuit
and from the RF wafer capacitance of the order of $\sim$ pF/gap. All
RF units are connected in parallel to the same RF source in our
setup. The typical pressure in the vacuum chamber is in the low
$\unit[10^{-6}]{Torr}$ range without gas flow and rises to
$\unit[2\times 10^{-5}]{Torr}$ when the gas flow is on. The power supplies as well
as the RF generator are remote controlled via
\citet{LabVIEW}. Furthermore, we monitor and control voltage and
current signals using National Instruments cDAQ modules. A Picoscope
(Picoscope 5443B) is used inside the high voltage rack that houses the
grid power supplies of the source to monitor voltages and currents of
the extraction system. The oscilloscope is controlled using
\citet{Python} running on a Raspberry Pi and communicates its data to
LabVIEW using \citet{ZMQ} over a fiber-coupled ethernet
connection. Timing is controlled by two Stanford Research Systems
DG535 modules which are connected to LabVIEW. A Tektronix Oscilloscope
is used to measure the Faraday cup signal, which is measured across a
\unit[10]{k\textohm} resistor.

A voltage scan on the retarding grid can be used to measure the energy
distribution of the beam. The long beam pulse of \unit[0.3]{ms} relative to
the RF frequency of \unit[15]{MHz} will create a continuous distribution of
ion energy where some particles are accelerated and others
decelerated. If the drift regions are designed for the right ion
species and velocities, some ions can achieve acceleration across all
gaps. Other ions arriving at the wrong phase of the RF signal will be
decelerated in the first acceleration gap and therefore be out of
phase in the following gaps. This leads to a distribution of
velocities that can be measured in our experiments. In the following
we describe a simple 1D-model to calculate the measured beam currents.

In a longer accelerator structure, the ions entering at the wrong
phase will be lost early on due to mismatched focusing and one will be
left with a beam of discrete packages (about 10\% of the beam
current). To achieve higher transportable currents a buncher section
can be integrated at the beginning of the accelerator to be
able to minimize the beam loss.

\section{1D-Model of the Retarding-Field Analyzer}
To interpret the ion current detected in the Faraday cup, we created a
simple model to describe our experiment. For this model we assume a
uniform ion beam profile of a given pulse length (long compared to the
RF period). We then discretize the beam into $N$ macro particle, evenly
spaced along the $z$ axis and assign the same initial velocity to each macro
particle. For each particle we track its energy, $z$-position, and
time. We model the acceleration gaps as zero-length that provide
instantaneous acceleration. Gap positions are calculated according to
a given RF frequency, phase, amplitude, ion mass, and velocity. We also
assume a single RF sine-wave with a fixed amplitude and frequency that
applies to all RF gaps with no phase shifts, reasonable for the signal
propagation time differences in the experiment. Then we trace each
particle through the system by calculating the time it takes to arrive
at the next gap, calculate the RF phase for that particle and
add/subtract the RF voltage at that phase to the particles energy. All
particles are tracked this way until they exit the last gap. At this
point, the transmitted beam current is simulated by simply recording
the macro-particles whose energy exceeds the grid voltage. The
particles are then traced to the Faraday cup position where we can
create a time histogram of the particles that can pass the retarding
grid. This histogram can be scaled to present the measured beam
current signal, as well as an average current. We implemented this
algorithm in Python.

\section{Experimental Results and Discussion}
\begin{figure}[t]\vspace*{4pt}
\centering
\includegraphics[width=\linewidth]{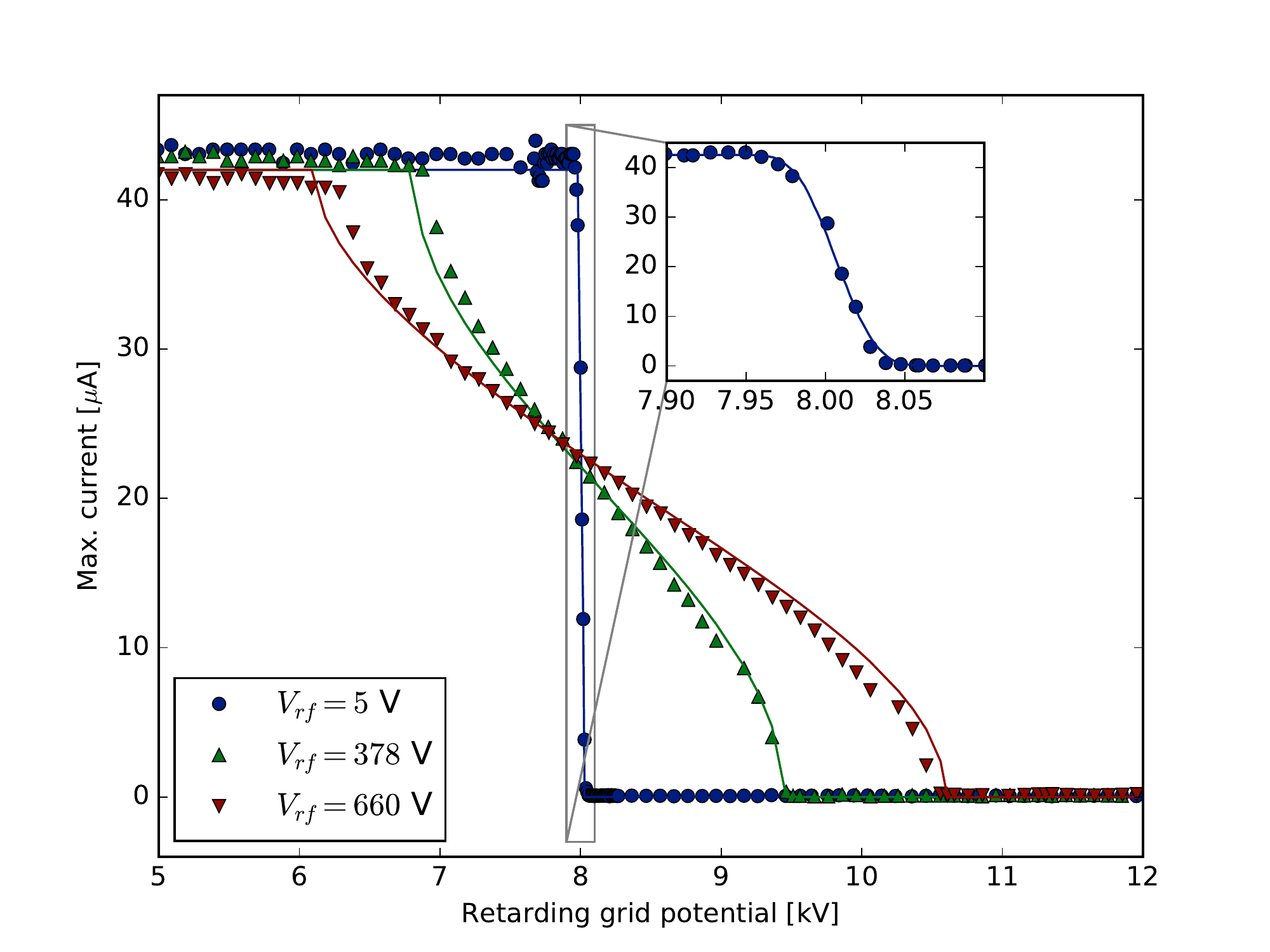}
\caption{Retarding-grid voltage scan with two RF units for low, medium
  and high RF voltages as described in the text. The inset shows a
  scan with the RF signal generator at its lowest setting. The
  simulations are shown as solid lines.}
\label{fig:2units}
\end{figure}

To demonstrate RF acceleration, we assembled two and three RF units
and scanned the retarding-grid voltage for different RF amplitudes. In
Fig.~\ref{fig:2units} we show the results of these measurements for
the case of two RF units consisting of four acceleration gaps. The
three measurements correspond to a high RF amplitude (\unit[300]{mV}
driving voltage at the RF-signal generator), a medium amplitude
(\unit[100]{mV}), and a low non-zero driving voltage
(\unit[1]{mV}). This latter case was chosen (vs. powering off the
electronics) to assure a well-defined voltage on the RF wafers for a
``no-RF case''.  The details of the resulting voltage scan are shown
in the inset of Fig.~\ref{fig:2units}. This measurement verifies our
incoming ion beam energy of \unit[8]{keV}.  Rather than a steep drop
to zero over \unit[0.5-2]{eV} (the expected plasma ion source
temperature) the current falls over \unit[10-15]{eV}. This apparent
energy spread is not an attribute of the beam but instead it is likely
due to the known lensing effect of the mesh and the associated
contribution to the intrinsic FWHM sensitivity of the diagnostic.  For
our mesh (90 lines/inch, $\unit[5.5 \times 10^{-3}]{inch}$ rectangular
mesh) the estimated FWHM energy resolution is \unit[15]{eV}, according
to the analytic model of \citet{Sakai1985}.

The solid lines in Fig.~\ref{fig:2units} show that simulation results
using our 1D-model are in good agreement with the experiments. Note
that the model has only one free parameters: the current level for low
grid voltage settings. All other parameters, such as the gap
distances, are taken from the experiment or measured directly, such as
the RF frequency and RF amplitude (measured with a high voltage
probe). To achieve a good match for the data shown in the inset of
Fig.~\ref{fig:2units}, we also assume a mesh resolution of
\unit[15]{eV}, which we implemented in the simulations by adding a
gaussian distribution to the starting energy of the ions. The effect
of an ion temperature would only be visible in the inset in
Fig.~\ref{fig:2units}, except that it is obscured by the \unit[15]{eV}
retarding-grid resolution. For the medium and high RF scans the ion
temperature has a negligible effect on the results. The simulations
therefore assume a monoenergetic beam.

\begin{figure}[t]\vspace*{4pt}
\centering
\includegraphics[width=\linewidth]{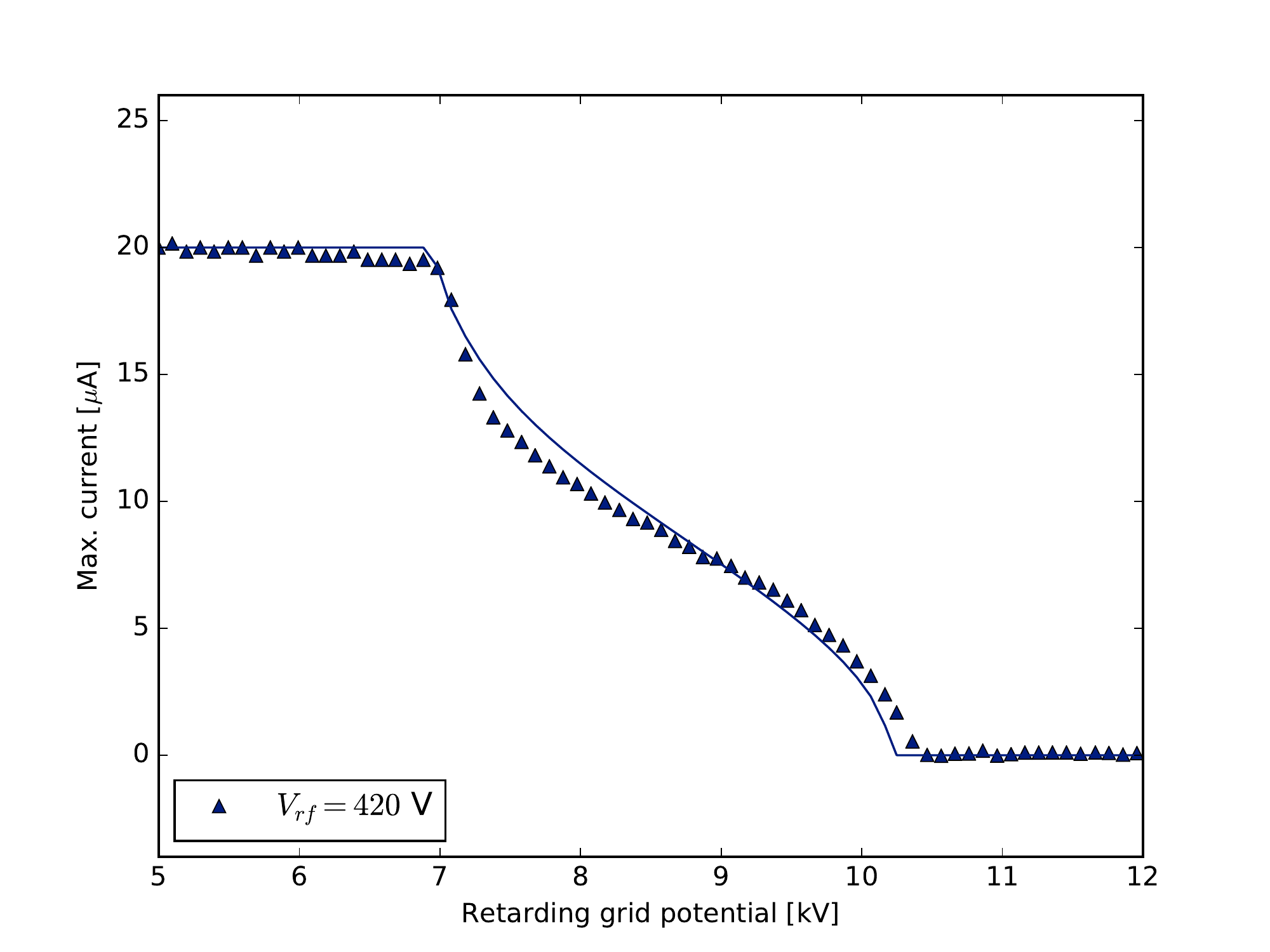}
\caption{Three RF units with an RF amplitude of \unit[420]{V} and a frequency
  of \unit[12]{MHz}.}
\label{fig:3units}
\end{figure}

In Fig.~\ref{fig:3units} results are shown from running three RF units (6
accelerations gaps). Since we are currently driving all RF gaps with a
single circuit, we had to lower the frequency to get the best resonant
condition in the circuit. The higher capacitance of the three RF unit
compared to the two RF units also resulted in a lower voltage per
wafer. Taking these effects into account, the results agree well with
our simulations.

The extracted beam current from the source for these experiments is in
both cases limited by the ion source conditions, which were optimized
for stable source operation instead of high beam currents.

\section{Outlook}
We showed the successful acceleration of ions using a compact
RF structure based on FR-4. Ongoing experiments are integrating ESQs
into the acceleration structure, as well as providing a six ESQ
matching section between the ion source and the MEQALAC structure to
be able to capture more ions from the source, reduce beam scraping and
establish matched beam conditions at the entrance of the
accelerator. In these experiments, the total acceleration is limited
by the voltage that the RF amplification can supply. As a result,
increasing the number of RF units did not appreciably increase the
total acceleration because the larger load caused the voltage per
acceleration gap to decrease. To be able to scale this technology to
higher output voltages, by using more RF units with a higher
acceleration gradient, we are working on integrating the RF generation
on the wafer using coplanar waveguides and also on-chip
RF generation. Using printed circuit boards enabled quick prototyping
of devices. However, using laser cutting technology and FR-4 limits us
to fabrication errors of $\unit[\sim 20]{\mu m}$. Fabricating the wafers in silicon
using standard optical lithography will reduce the fabrication error
by another factor of 10. Allowing smaller beam apertures and therefore
higher ion beam current. Lithography will also allow us to use large
parallel arrays of beamlets and therefore increase the transported
current.

We envision this technology to be applicable to beam energies between
\unit[100]{keV} to several \unit{MeV}. Below \unit[100]{keV} a single
vacuum gap to accelerate the ions can be used easily and above several
\unit{MeV} the length of the unit will be a limiting factor unless one
also increases the RF frequency. As for the beam current densities we
estimate around \unitfrac[1]{mA}{cm$^2$} (when averaged over the total
cross section of the accelerator structure of e.g. a 4" wafer).  The
overall efficiency of a MEQALAC structure scales favorable compared to
other technologies, a comparison with radio-frequency quadrupole
linear accelerators can be found in paper of Urbanus, \textit{et al.}
\cite{Urbanus1990}.  Applications areas range from material analysis
to fusion drivers.

\section*{Acknowledgements}
We are grateful for insightful discussions with Andris Faltens
(LBNL). This work was supported by the Office of Science of the US
Department of Energy through the ARPA-E ALPHA program under contract
DE-AC02–05CH11231 (LBNL).

\bibliography{paper}
\bibliographystyle{elsarticle-harv}

\end{document}